\documentstyle[12pt]{article}
\begin{document}
\thispagestyle{empty}

\begin{center}
\LARGE \tt \bf{Riemannian geometry of turbulent gravity wave  black hole analogs}
\end{center}

\vspace{2.5cm}

\begin{center} {\large L.C. Garcia de Andrade \footnote{Departamento de
F\'{\i}sica Te\'{o}rica - Instituto de F\'{\i}sica Rua S\~{a}o Fco. Xavier 524, Rio de Janeiro, RJ

Maracan\~{a}, CEP:20550-003 , Brasil.

E-Mail.: garcia@dft.if.uerj.br}}
\end{center}

\vspace{2.0cm}

\begin{abstract}
The gravity water wave black (GWBH) hole analog discovered by Schutzhold and Unruh (SU) is extended to allow for the presence of turbulent shear flow. The Riemannian geometry of turbulent black holes (BH) analogs in water waves is computed in the case of laminar tirbulent shear flow. The Riemann curvature is constant and the geodesic deviation equation shows that the curvature acts locally as a diverging lens and the stream lines on opposite sides of the analog black hole flow apart from each other. In this case it is shown that  the curvature quantities can be  expressed in terms of the Newtonian gravitational constant in the ergoregion. The dispersion relation is obtained for the case of constant flow injection.
\end{abstract}

\newpage

\section{Introduction}

Schutzhold and Unruh \cite{1} have built water gravity wave analogs of black holes. Since the gravity water waves speed can be adjusted easily by varying the height of the fluid, these scenarios possess certain advantages over the more well known dielectric and acoustic black holes analogs. In acoustic BHs the acoustic horizon appears if the velocity of the fluid exceeds the speed of sound in the medium of the liquid. The  Riemannian geometry of vortex acoustics in such vortex sinks \cite{2} has been investigated in detailed by Fischer and Visser \cite{3,4}. Recently extension of this geometry to the non-Riemannian geometry of vortex acoustics to taking into acount rotating non-viscous and rotating viscous fluids have been discussed by Garcia de Andrade \cite{5,6}. In this paper we investigate the detailed geometry of the turbulent GWBH effective metric. As an example we apply  Riemannian geometry to the turbulent acoustic shear flow recently investigated by J. de Barros in her DSc. thesis \cite{7}. Her model is composed of shear laminar flows in between two other fluid layers.  The shear layers here represents the surface of the water from above and the bottom of the water tank on the other side of the  fluid. The paper is organised as follows: In the section two we present the physics of the model \cite{7} as well the metric of the turbulent fluid shear layer endowed with an  ergoregion of the analog GWBH. In section 3 the computation of Christoffel connections and Riemann and Einstein tensors are given and the application to a shear laminar turbulent flow is given which allows us to compute the curvature quantities in terms of the Newtonian gravitational constant. In section four the computation of the dispersive relation of the turbulent irrotationally perturbed flow in the case of stationary injection of mass into the fluid, and the group velocity of the flow. Finally in the conclusions we discuss applications and future prospects. 
\section{The turbulent fluid model and the water GWBH}
Turbulent phenomena are in general quite complicated and can only be treated numerically and seldom treated analytically \cite{8}. Nevertheless some interesting particular models of shear flow turbulence can be treated analytically such as the case of laminar shear flows in between two layers where the background turbulence is perturbed. In this section we shall consider one of these models described by J.D. Barros \cite{8} and show that they can be given a geometrical interpretation as the Riemannian geometry of a GWBH. The dynamical equations are given by the modified Euler and conservation equations 
\begin{equation}
\frac{D{\rho}}{Dt}+{\rho}{\nabla}.\vec{v}=q(t)
\label{1}
\end{equation}
\begin{equation}
{\rho}\frac{D{\vec{v}}}{Dt}+{\nabla}p= \vec{f}
\label{2}
\end{equation}
where the source $\vec{f}$ is a dipolar source while the turbulence injection is given by the function $q(t)$. The differential operator $\frac{D}{Dt}$ is given by 
\begin{equation}
\frac{D_{1}}{Dt}=(\frac{\partial}{{\partial}t}+\vec{U}(x,y).{\nabla})
\label{3}
\end{equation}
Linearization of the dynamical equations of the flow according to the scheme
\begin{equation}
{\rho}={\rho}_{0}+{\rho}'
\label{4}
\end{equation}
\begin{equation}
p=p_{0}+ p'
\label{5}
\end{equation}
\begin{equation}
\vec{v}=\vec{U}+\vec{v'}
\label{6}
\end{equation}
where ${\nabla}{\rho}_{0}=0={\nabla}p_{0}$, where $c_{0}$ is the speed of sound in the medium and $\vec{U}=U(x,y)\vec{e_{1}}$ is the velocity of the background turbulent flow, yields the following dynamical flow equations
\begin{equation}
\frac{D{\rho}'}{Dt}+{\rho}_{0}{\nabla}.\vec{v'}=q
\label{7}
\end{equation}
\begin{equation}
{\rho}_{0}(\frac{D_{1}\vec{v'}}{Dt})+({\nabla}U.\vec{v'})\vec{e_{1}}+{\nabla}p'= \vec{f}
\label{8}
\end{equation}
To simplify matters we shall assume that the flow is irrotational or $\vec{v'}={\nabla}{\psi}$ where ${\psi}$ is a scalar function. This assumption is reasonable since we shall next consider a shear thin layer  where no bulk vorticity exists and only localized vortices are appreciable. We also assume that the external force can be obtained by the relation $\vec{f}= {\nabla}{\phi}$ and ${\nabla}U.\vec{v'}=0$. This term can be incorporated again when we investigate the non-Riemannian geometry of turbulent flows \cite{3} later on. The external force here is similar to what happens in the Newtonian gravitational analogue of gravity waves \cite{4}. After some usual algebraic manipulations one obtains
\begin{equation}
{\nabla}^{2}{\psi}-\frac{1}{{c^{2}}}\frac{{D_{1}}^{2}{\psi}}{D{t}^{2}}=\frac{1}{{\rho}_{0}}[q-\frac{D_{1}}{Dt}(\frac{{\phi}}{{c}^{2}})]
\label{9}
\end{equation}
By making use of expression (\ref{3}) into (\ref{1}) yields the generalised wave equation for the turbulent flow
\begin{equation}
c^{2}{\nabla}^{2}{\psi}-(\frac{{\partial}}{{\partial}{t}}+\vec{U}.{\nabla})^{2}{\psi}=\frac{c^{2}}{{\rho}_{0}}[q-\frac{D_{1}}{Dt}(\frac{{\phi}}{{c}^{2}})]
\label{10}
\end{equation}
Throughout the calculations we also use the following relation
\begin{equation}
\frac{D_{1}{\rho}'}{Dt}= \frac{1}{{c^{2}}}\frac{D_{1}p'}{Dt}
\label{11}
\end{equation}
This covariant scalar wave equation (\ref{10}) gives rise to the following effective metric 
\begin{equation}
{g^{00}}_{eff}=1
\label{12}
\end{equation}
\begin{equation}
{g^{0i}}_{eff}= U^{i}
\label{13}
\end{equation}
and
\begin{equation}
{g^{ij}}_{eff}= U^{i}U^{j}-c^{2}{\delta}^{ij}
\label{14}
\end{equation}
where ${\delta}^{ij}$ is the Kronecker delta and $(i,j=1,2,3)$. The line element representing the analog effective metric is
\begin{equation}
ds^{2}=-(1-\frac{{\vec{U}}^{2}}{c^{2}})dt^{2}-2U_{i}dx^{i}dt+ {\delta}_{ij}dx^{i}dx^{j}
\label{15}
\end{equation}
This "spacetime" region represents the PGL type metric of the analog BH which reduces to the Schutzhold-Unruh metric of gravity water waves when $c^{2}=gh$ where g is the usual Newtonian gravity acceleration and h is the height of the  tank of water. From the above metric one notes that the ergoregion of the effective BH is given by ${\vec{U}}^{2}=c^{2}$.
\section{Riemannian curvature of the GWBH and gravitational analog of diverging lens}
In this section we shall compute following reference \cite{9} the curvature quantities associated with the GWBH. We also compute the geodesic deviation equation and show that the GWBH curvature acts locally as a diverging lens with a complex curvature frequency. Let us now consider the Riemann curvature components by applying the  Gauss-Codazzi decomposition used by Fisher and Visser in reference \cite{9}. Applying their expressions to our particular example one obtains, first the Riemann-Christoffel connections
\begin{equation}
{{\Gamma}^{t}}_{ij}=D_{ij}=({\partial}_{i}U_{j}+{\partial}_{j}U_{i})
\label{16}
\end{equation} 
\begin{equation}
{{\Gamma}^{t}}_{tt}= U_{i}U_{j}D_{ij}=\frac{1}{2}(\vec{U}.{\nabla}){\vec{U}}^{2}
\label{17}
\end{equation} 
\begin{equation}
{{\Gamma}^{t}}_{it}=- U_{j}D_{ij}
\label{18}
\end{equation} 
\begin{equation}
{{\Gamma}^{t}}_{it}= - U_{j}D_{ij}
\label{19}
\end{equation}  
\begin{equation}
{{\Gamma}^{i}}_{jk}= U_{i}D_{jk}
\label{20}
\end{equation}
\begin{equation}
{{\Gamma}^{i}}_{tt} = -{\partial}_{t}U_{i}-\frac{1}{2}({\delta}_{ij}-U^{i}U^{j}){\partial}_{j}{\vec{U}}^{2}
\label{21}
\end{equation}
where the first term vanishes since $U^{i}$ does not depend on time  
\begin{equation}
{{\Gamma}^{i}}_{tj} = -U^{i}U^{k}D_{jk}
\label{22}
\end{equation}
where we have put the vorticity ${\Omega}_{ij}$ to vanish since there is no vorticity in the Lily model considered here and the perturbation $\vec{v'}$ is irrotational. With these tools in hand we are able to compute the Riemann curvature components for the model where we consider the $\vec{U}= U(x,y){\vec{e}}_{1}$ where $U=By$ , B being a proportionality constant, one obtains the following curvature components
\begin{equation}
R_{ijkl} =(D_{kl}D_{ij}-D_{jl}D_{ik})
\label{23}
\end{equation}
\begin{equation}
R_{tijk} = U_{l}(D_{kl}D_{ij}-D_{jl}D_{ik})
\label{24}
\end{equation}
\begin{equation}
R_{titj} = -(D^{2})_{ij}- U_{k}U_{k,ij}+ U_{k}U_{l}(D_{kl}D_{ij}-D_{jk}D_{il})
\label{25}
\end{equation}
where we have used also the expression $TrD=div\vec{U}=B$. To simplify the computations we consider the Riemann components in a orthonormal tetrad Minkowski frame $(e^{a}_{\mu})$, where ${\mu}=0,1,2,3$. The metric in the tetrad frame is ${\eta}_{ab}$ where $(a=0,1,2,3)$, is given in terms of coordinate Riemann metric as   
\begin{equation}
g_{{\mu}{\nu}} = {\eta}_{ab}{e^{a}}_{\mu}{e^{b}}_{\nu}
\label{26}
\end{equation}
where we use the same simple gauge used by Fischer and Visser \cite{9} as ${e^{t'}}_{t}=-1$, ${e^{i'}}_{t}=-v^{i}$ and ${e^{j'}}_{i}= {{\delta}^{j'}}_{i}$. Finally the Riemann components in this gauge are
\begin{equation}
R_{t'y't'y'} = -D_{yx}D_{xy}=-B^{2}
\label{27}
\end{equation}
\begin{equation}
R_{x'y'x'y'} = -{D^{2}}_{yx}=-B^{2}
\label{28}
\end{equation}
These spacetime reveal that this particular gravity wave black hole analog with laminar shear flow has a constant Riemann curvature. In this same gauge the Ricci components are
\begin{equation}
R_{t'i'} = -{\nabla}^{2}U_{i}= -\frac{1}{2}({\nabla}{\times}\vec{\omega})_{i}=0
\label{29}
\end{equation} 
and the only non-vanishing components of the Einstein tensor are
\begin{equation}
G_{t't'} = -\frac{1}{2}B^{2}
\label{30}
\end{equation} 
\begin{equation}
G_{x'y'} = B^{2}
\label{31}
\end{equation} 
With these curvature components we are now able to calculate the geodesic deviation equation
\begin{equation}
\frac{D^{2}}{d{\lambda}^{2}}[{\delta}y]+[(R_{y't'y't'}+R_{y'x'y'x'})(u^{t'})^{2}][{\delta}y]=0
\label{32}
\end{equation}   
where ${\delta}y$ is a space-like separation of a family of geodesics in the $x-direction$ with tangent vector ${\vec{u^{a}}}=(u^{t'},u^{t'},0,0)$ and $\vec{n}=(0,0,{\delta}y,0)$ is the separation vector of geodesic deviation. Substitution of the curvature expressions (\ref{26}) and (\ref{27}) into equation (\ref{32}) yields
\begin{equation}
\frac{D^{2}}{d{\lambda}^{2}}[{\delta}y]-2B^{2}(u^{t'})^{2}[{\delta}y]=0
\label{33}
\end{equation}
Since the geodesic deviation equation can be considered as a parametrically driven harmonic oscillator \cite{9} in the affine parameter ${\lambda}$ with frequency   
\begin{equation}
{\Omega}(t)= u^{t'}\sqrt{R_{y't'y't'}+R_{y'x'y'x'}}(u^{t'})
\label{34}
\end{equation}
in the turbulent shear layer considered here the frquency is then imaginary or ${\Omega}({\lambda})= iB$ which leads us to the physical conclusion that the Riemann curvature acts locally as a diverging lens in the effective spacetime of the GWBH.     
\section{Dispersion relation for the turbulent BH analog}
By making use of the ansatz  
\begin{equation}
{\psi}(t,\vec{r})={\psi}_{0}(\vec{r})e^{-i({\omega}t-\vec{k}.\vec{r})}
\label{35}
\end{equation}
and
\begin{equation}
{\phi}(t,\vec{r})={\phi}_{0}(\vec{r})e^{-i({\omega}_{0}t-\vec{k_{0}}.\vec{r})}
\label{36}
\end{equation}
and substituting these expressions into the equation (\ref{11}) one obtains the following dispersion relation
\begin{equation}
(c^{2}{\vec{k}}^{2}-{\omega}^{2})-[\vec{U}.{\nabla}]{\beta}-{\beta}^{2}-2i{\omega}{\beta}= \frac{c^{2}}{{\rho}_{0}{\psi}}[q(t,\vec{r})-i\frac{{\omega}}{c^{2}}{\phi}-{\beta}\frac{{\phi}}{c^{2}}]
\label{37}
\end{equation}
where ${\beta}=\vec{U}.\vec{k}$. From expression (\ref{18}) we  note that there are complex terms in frequency ${\omega}$ which implies that we have dissipation terms in the dispersion relation. To simplify matters we make use of the ressonance frequency ${\omega}={\omega}_{0}$ and $\vec{k_{0}}=\vec{k}$ and further assuming that the amplitude of the force ${\phi}_{0}$ is much smaller than the amplitute of the potential flow ${\psi}_{0}$ one obtains the final dispersion relation
\begin{equation}
{{\omega}_{0}}^{2}=c^{2}k^{2}-[\vec{U}.{\nabla}]{\beta}-{\beta}-\frac{c^{2}}{{\rho}_{0}}\frac{q_{0}}{{\psi}_{0}}
\label{38}
\end{equation}
where $q_{0}$ is the amplitude of the constante injection of the generator of turbulence. Note also that there is no dissipation in this dispersion relation. From the more general expression (\ref{18}) one notes that the Lorentz acoustic type violation is also present here dueto the turbulent terms. A simple algebraic manipulation in these formulas allow us to determine a simple expression for the group velocity
\begin{equation}
v_{g}=\frac{d{\omega}}{dk}=\frac{1}{2}[1-q_{0}]v-i{\vec{U}}.\vec{k}
\label{39}
\end{equation}
where $v:=\frac{c^{2}}{v_{p}}$ where $v_{p}$ is the phase velocity. The frequency is given by
\begin{equation}
{\omega}=\frac{c^{2}k}{v_{g}}[1-q_{0}]
\label{40}
\end{equation}
Here we used the approximation $|{\vec{U}}|<< v_{g}$ \cite{10} which was used by Berry et al in the investigation of the wavefronts dislocations in the Aharonov-Bohm effect in its water wave analog.
\section{conclusions}
A Riemannian shear flow thin layer is shown to yield an effective metric for a non-rotating black hole in water turbulent waves. The idea presented here extends Schutzhold-Unruh gravity water waves BH. Non-Riemannian turbulent flows and their respective BH analogs can also be generated if one considers the vorticity of the fluid besides the shear considered here. This work is under progress. Another interesting future prospect is the investigation of the AB effect around the GWBH as was done around vortices by F. Lund \cite{11}. Actually since in these papers dislocation is treated in water a Cartan torsion vector could be associated to a non-Riemannian geometry of turbulent shear flow like a black hole analog in water waves. In this respect knowledge of the non-Riemannian structure of analog models in vortex fluid mechanics and acoustics \cite{5} would be very  useful.
\section*{Acknowledgement}
I thank P.S.Letelier and Ivano D. Soares for helpful discussions on the subject of this paper, and to CNPq. and UERJ for financial support. 

\end{document}